\documentclass[useAMS,usenatbib]{mn2e}
\usepackage{times}
\usepackage{graphicx}

\usepackage[T1]{fontenc}
\usepackage{aecompl}

\newcommand\lsim{\mathrel{\rlap{\lower4pt\hbox{\hskip1pt$\sim$}}
        \raise1pt\hbox{$<$}}}
\newcommand\gsim{\mathrel{\rlap{\lower4pt\hbox{\hskip1pt$\sim$}}
        \raise1pt\hbox{$>$}}}

\usepackage{graphicx}
\usepackage{verbatim}

\usepackage{calligra}
\DeclareMathAlphabet{\mathcalligra}{T1}{calligra}{m}{n}
\DeclareFontShape{T1}{calligra}{m}{n}{<->s*[2.2]callig15}{}

\def\bin{\rm{bin}}
\def\obs{\rm{obs}}
\def\var{\rm{var}}

\def\Msun{ \rm M_{ \odot } }

\date{Accepted 2015 June 30. Received 2015 June 26; in original form 2015 February 11}

\volume{452} \pagerange{2540--2545} \pubyear{2015}

\begin{document}

\title[A Reduced Orbital Period for PG 1302-102]{A reduced orbital period for the supermassive black hole binary candidate in the quasar PG 1302-102?}

\author[D. J. D'Orazio et al.]{D. J. D'Orazio$^1$\thanks{dorazio@astro.columbia.edu}, Z.~Haiman$^1$,  P. Duffell$^2$, B. D. Farris$^{1,3}$ and A. I. MacFadyen$^3$ \\ \\
$^1$Department of Astronomy, Columbia University, 550 West 120th Street, New York, NY 10027, USA\\
$^2$Theoretical Astrophysics Center, Department of Astronomy, University of California, Berkeley, CA 94720, USA\\
$^3$Center for Cosmology and Particle Physics, Physics Department, New York University, New York, NY 10003, USA}

\maketitle

\begin{abstract}
Graham et al. have detected a 5.2 yr periodic optical
variability of the quasar PG~1302-102 at redshift $z=0.3$, which they
interpret as the redshifted orbital period $(1+z)t_{\rm bin}$ of a
putative supermassive black hole binary (SMBHB). Here, we consider the
implications of a $3-8$ times shorter orbital period, suggested by
hydrodynamical simulations of circumbinary discs (CBDs) with nearly
equal--mass SMBHBs ($q\equiv M_2/M_1\gsim 0.3$).  With the
corresponding $2-4$ times tighter binary separation, PG~1302 would be
undergoing gravitational wave dominated inspiral, and serve as a proof
that the BHs can be fuelled and produce bright emission even in
this late stage of the merger. The expected fraction of binaries with
the shorter $t_{\rm bin}$, among bright quasars, would be reduced by one to two
orders of magnitude, compared to the 5.2 yr period, in better
agreement with the rarity of candidates reported by
Graham et al.  Finally, shorter periods would imply higher
binary speeds, possibly imprinting periodicity on the light curves
from relativistic beaming, as well as measurable relativistic effects
on the Fe K $\alpha$ line.  The CBD model predicts additional periodic
variability on time-scales of $t_{\rm bin}$ and $\approx 0.5 t_{\rm
  bin}$, as well as periodic variation of broad line widths and
offsets relative to the narrow lines, which are consistent with
the observations.  Future observations will be able to test these
predictions and hence the binary+CBD hypothesis for PG~1302.
\end{abstract}


\section{Introduction}
\citet[][hereafter G15]{Graham:2015} recently reported strong optical
variability of the quasar PG 1302-102, with an observed period of
$t_{\obs} = 5.2 \pm 0.2$ yr. G15 attribute the variability to the
orbital motion of a super-massive black hole binary (SMBHB). Broad
emission lines in the spectrum of PG~1302 imply a binary mass in the
range $M = 10^{8.3-9.4} \Msun$. Assuming that the binary's orbital
period $t_{\rm bin}$ equals the rest--frame optical variability period
$t_{\rm opt}$, G15 derive a fiducial binary separation $a\approx
(0.0084 \pm 0.0003)\mbox{pc} \approx (276 \pm 9) R_{\rm{S}}$ for $M=10^{8.5}
\Msun$, where $R_{\rm{S}} = 2GM/c^2$ is the Schwarzschild radius.

Hydrodynamical simulations of a binary BH embedded in a gaseous
accretion disc predict that, depending on the binary mass ratio and
the physical parameters of the disc, the strongest periodicity in the
accretion rate on to the BHs may correspond to the motion of gas
farther out in the disc, at a few times the binary separation,
producing optical variability at several ($\sim$3-8) times the binary
orbital period. In this article, we discuss this expectation, and
show that a reduced binary period, in the case of PG 1302, would have
several important implications. Follow up spectroscopy and photometric
monitoring can determine the true binary period.

The rich variability structure of the mass accretion rates seen in
simulations can be roughly divided into four distinct categories,
based on the binary mass ratio $q\equiv M_2/M_1$. For $q\lsim0.05$,
the disc is steady and the BH accretion rate displays no strong
variability \citep[][D'Orazio et al. in preparation]{DHM:2013:MNRAS, Farris:2014}. 
For $0.05 \lsim q \lsim 0.3$, the accretion rate
varies periodically on the time-scale $t_{\rm bin}$, with additional
periodicity at $\approx 0.5t_{\rm bin}$. Binaries with $0.3 \lsim q
\lsim 0.8$ clear a lopsided central cavity in the disc, causing
variability on three time-scales. The dominant period, $(3-8) t_{\rm
  bin}$ is that of an over dense lump, orbiting at the ridge of the
cavity, with additional periodicities at $t_{\rm bin}$ and $\approx
0.5t_{\rm bin}$ \citep{MacFadyen:2008, ShiKrolik:2012:ApJ, Noble+2012,
  Roedig:2012:Trqs:arxiv, DHM:2013:MNRAS, Farris:2014}.  The dominant
period depends on the size of the cavity, and thus on disc parameters,
such as temperature and viscosity. Finally, equal-mass ($q=1$)
binaries display variability at the longer lump period and at $\approx
0.5t_{\rm bin}$.

Here, we consider the identification of the observed variability of
PG~1302 with the long, cavity-wall period, and introduce the parameter
$\chi\equiv t_{\rm opt}/t_{\bin}$, denoting the ratio,
$3\lsim\chi\lsim8$, of the observed rest-frame period and the true
binary period.  The binary separation is then
\begin{equation}
a \approx (94\pm3) R_{\rm{S}}  \left(\frac{\chi}{5}\right) ^{-2/3}  \left(\frac{M}{10^{8.5}\Msun}\right)^{-2/3},
\label{Eq:a*}
\end{equation}
or $(0.0029 \pm 0.0001)$ pc for the fiducial choices of $\chi$ and
$M$. 

In the rest of this article, we first (\S\ref{S:Implications})
explore several implications of a reduction in the binary's orbital
period, including the nature of PG~1302's orbital decay and its
ability to produce electromagnetic (EM) radiation
(\S\ref{SS:Decoupling}), the expected binary fraction of quasars
(\S\ref{SS:Rates}), and the detectability of gravitational waves (GWs)
from PG~1302 by pulsar timing arrays (PTAs).  We then
(\S\ref{S:Predictions}) propose possible observational tests of the
underlying binary BH + circumbinary disc (CBD) model, including variations
of broad line widths and centroids correlating with the optical
variability (\S \ref{SS:BLs}), additional periodic variability at the
true $t_{\rm bin}$ caused by relativistic beaming (\S\ref{SS:Rel}),
signatures in the broad Fe~K $\alpha$ lines (\S\ref{SS:Rel}), and the
existence of distinct secondary peaks in the periodogram
(\S\ref{SS:OTV}).  We briefly summarize our main conclusions in
\S\ref{S:Conclude}.

\begin{figure}
\begin{center}$
\begin{array}{cc}  \hspace{-10 pt} 
\includegraphics[scale=0.367]{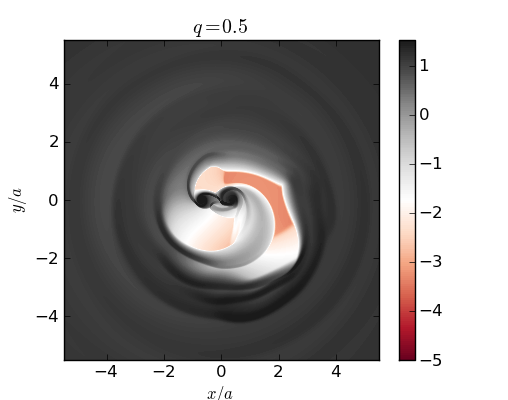} &    \hspace{-25 pt} 
\includegraphics[scale=0.245]{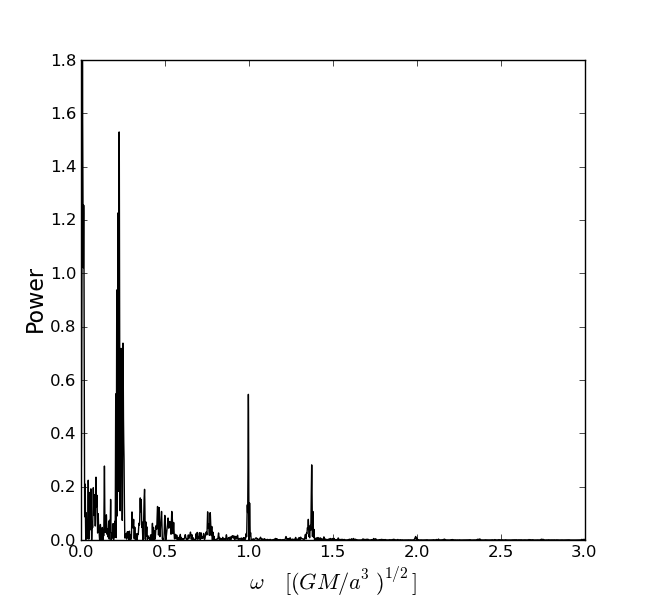} \\
 \hspace{-10 pt}
  \includegraphics[scale=0.367]{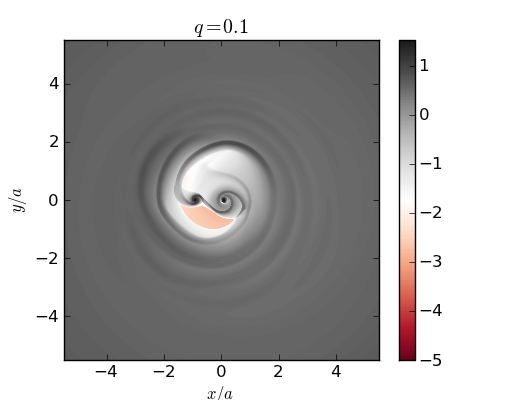} &    \hspace{-25 pt}
\includegraphics[scale=0.245]{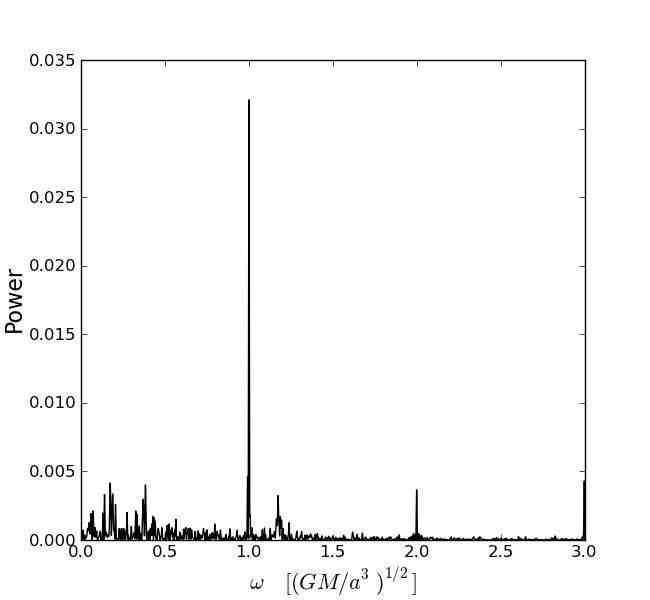}
  \vspace{-10 pt}
\end{array}$
\end{center}
\caption{Results of 2D hydrodynamical simulations of a binary BH
  surrounded by a circumbinary accretion disc.  The BHs clear out a
  central cavity and form their own minidiscs. {\em Left-hand panels:}
  snapshots of the (logarithmic) surface-density of the gas discs,
  after reaching quasi-steady state, with mass ratios of $q=0.5$ (top)
  and $q=0.1$ (bottom). {\em Right-hand panels:} corresponding LSPs 
  of the total accretion rate on to the secondary + primary BHs. The discs 
  are locally isothermal with a Mach number of 10 and an alpha viscosity 
  prescription ($\alpha=0.1$).}
\label{Fig:DensLS}
\end{figure}

\vspace{-\baselineskip}
\section{Implications of a shorter orbital period}
\label{S:Implications}

In order to demonstrate the possibility of a short orbital period for
the PG 1302 binary, we have performed hydrodynamical simulations,
following the set-up in our earlier work~\citep{Farris:2014}.  The
hydrodynamical equations are evolved for $\gsim 600$ binary orbits, 
using the two-dimensional code DISCO \citep{Duffell:2011:TESS}, with the BHs moving
on fixed circular orbits, surrounded by an isothermal (Mach number =
10) disc, obeying an $\alpha$-viscosity prescription
($\alpha=0.1$). The fluid motion around individual BHs is well
resolved (with a log grid of $384$ radial cells extending to $8a$ and a maximum of $512$ azimuthal cells).  
The two runs discussed below differ only in their BH mass ratio ($q=0.1$ and
$q=0.5$).  A wider range of simulations is needed in the future, to
address possibilities such as
eccentric~\citep{Roedig:2012:Trqs:arxiv},
tilted~\citep{Hayasaki+2015}, or
retrograde~\citep{Nixon:2011:LongSim} binary orbits.

The results are illustrated in Fig \ref{Fig:DensLS}. The left-hand panels
show snapshots of the surface density, and the right-hand panels show
Lomb Scargle periodograms (LSPs) of the total accretion rate measured
in the two BH minidiscs over 200 binary orbits. The top panels, for
$q=0.5$, show an over dense lump orbiting at the rim of the central
cavity, resulting in strong periodicity at the orbital time $\approx
6t_{\rm bin}$ of the lump. The periodogram shows weaker peaks at
$t_{\bin}$ and at $\sim 0.6t_{\bin}$.  This three-time-scale behaviour,
with the longest time-scale dominating, is observed for $0.3 \lsim q
\lsim 0.8$.  The location of the highest frequency peak is closer to $0.5
t_{\bin}$ near the low end of this range ($q\sim 0.3$), and also has a
weak dependence on disc temperature and viscosity which must be quantified in future work.  The bottom panels, for $q=0.1$,
show no orbiting lump and exhibit accretion rate periodicity only at
$t_{\bin}$ and $0.5t_{\bin}$.  This behaviour is found in the range
$0.05 \lsim q \lsim 0.3$.

\cite{Farris:2014} have shown that for unequal-mass binaries,
accretion occurs preferentially on to the secondary BH, with the ratio
of accretion rates as skewed as $\dot{M_2}/\dot{M_1}\approx 10-20$ in
the range $0.02\lsim q \lsim 0.1$.  Over long time-scales, this would
drive the binary to more equal masses, suggesting that mass ratios of
$0.3 \lsim q \lsim 0.8$ may be common.  Near-equal mass binaries are
also preferred in cosmological models of the population of merging
SMBHs \citep{Volonteri:2003}.

Although there are large uncertainties in how accretion rate fluctuations turn
into luminosity variations, we do expect the simulated accretion rate variations to lead to optical luminosity variations for PG~1302. The luminosity will follow local accretion rate fluctuations when the longer of the thermal or photon diffusion time-scale is much shorter than the accretion rate fluctuation time-scale ($\sim t_{\rm bin}$). This is indeed the case where accretion modulations occur in our simulations, at the minidisc edges. Furthermore, optical emission is generated at the minidiscs edges. We compute thin disc spectra for the CBD and minidiscs. For the preferred mass range of PG 1302, near unity mass ratios, and the expected range of $\chi$, the dominant optical component of the spectrum is generated by the low-energy tail of the blackbody emission from the outer edges of the minidiscs, as well as (steady) emission from the inner regions of the CBD. Disc+binary simulations by \cite{Farris:2015:Cool}, which self-consistently compute the local effective disc temperature not assuming a steady state, find results in agreement with our analytic reasoning: luminosity variations track the accretion-rate fluctuations, except at low frequencies where the quasi-steady CBD dominates.

The above lines of evidence motivate us to examine the possibility that the
apparent $5.2$ yr period in PG 1302 is the (redshifted) lump period,
and assess the implications.

\subsection{Binary-Disc Decoupling}
\label{SS:Decoupling}

\begin{figure}
\begin{center}
\includegraphics[scale=0.36]{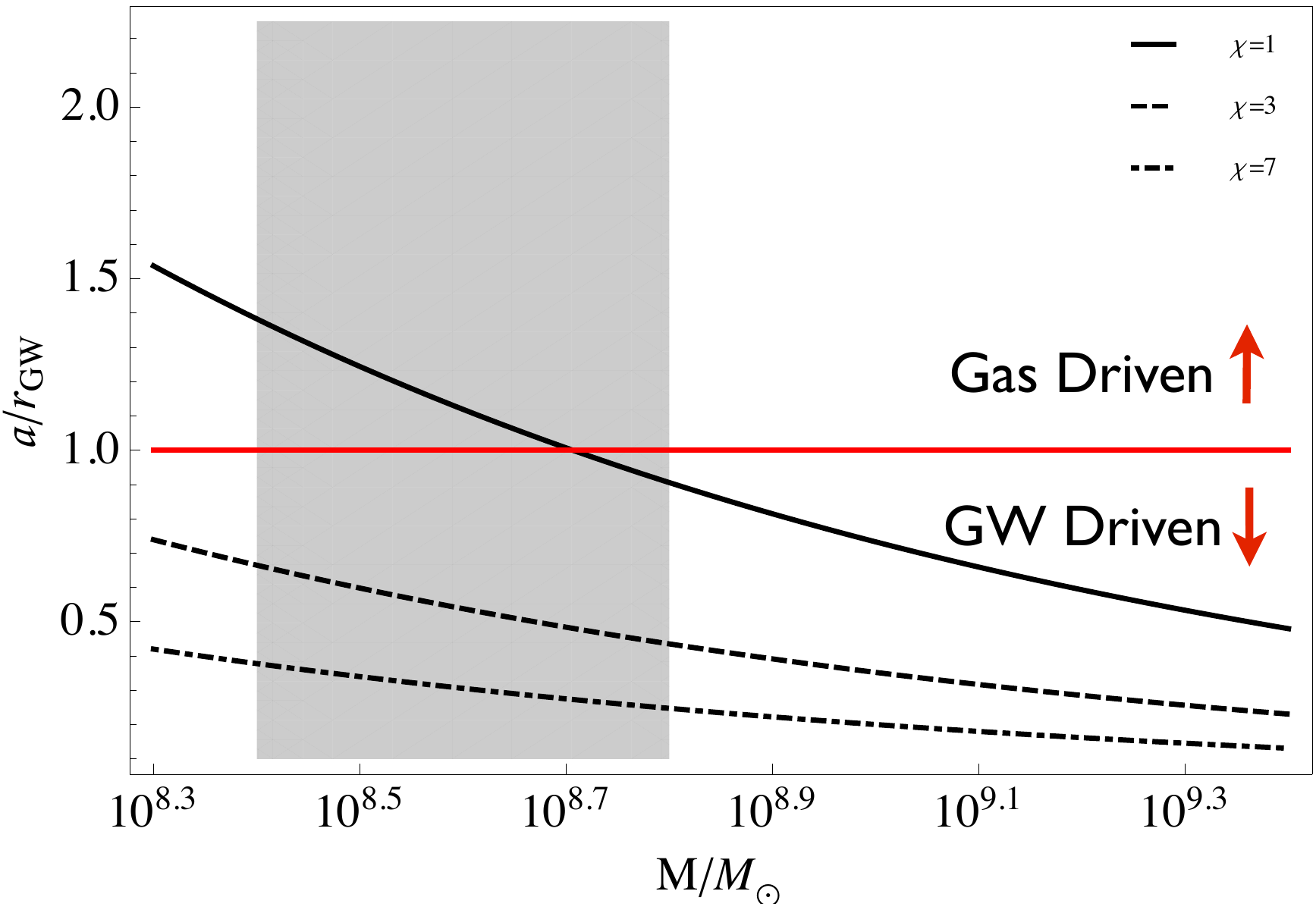}  \vspace{-15 pt}
\end{center}
\caption{The ratio of binary separation $a$ to the decoupling radius
  $r_{\rm GW}$, for three different values of the ratio between the
  rest frame optical period and the true binary period, $\chi=t_{\rm
    opt}/t_{\bin}$. The shaded region marks the binary mass
  range inferred from the widths of broad lines measured by G15. For $\chi>3$,
  the PG~1302 binary is past decoupling, for any choice of mass $M$.}
\label{Fig:Decoupling}
\end{figure}

A shorter orbital period would place the binary at a later stage of
its orbital decay.  A critical point during the orbital decay is the
decoupling of the binary from the CBD, and it is important to know
whether the binary is past this point. 
Here, we consider the decoupling radius for which the GW decay time-scale becomes shorter than the decay time-scale due to gaseous torques (so-called secondary-dominated Type II migration; \citealt{SyerClarke95}), outpacing the CBD.
We use simple 1D models of the binary
+ disc system \citep[][hereafter HKM09]{HKM09} to calculate the
separation $r_{\rm GW}$ at which decoupling occurs for a circular
binary with mass ratio $q=0.3$. We assume an $\alpha$-viscosity $\nu =
\alpha P_{\rm{gas}} (\rho\Omega)^{-1}$, with gas pressure
$P_{\rm{gas}}$, density $\rho$, and disc angular velocity $\Omega$. All other disc parameters are assumed to have the
fiducial values given in HKM09.  In Fig. \ref{Fig:Decoupling}, we plot
the ratio $a/r_{\rm{GW}}$ as a function of the total binary mass $M$, with the binary separation from equation~(\ref{Eq:a*}), for
the range of masses in G15, and for three values of $\chi$ covering
the range suggested by the hydrodynamical simulations.

Interpreting the observed variability in PG 1302 with $t_{\rm bin}$, as
may be justified for $q\lsim 0.3$, it is unclear whether or not the
binary has entered the GW dominated regime and decoupled from the
disc. The binary would still be coupled to the disc for 
$M<10^{8.7} {\Msun}$ (the majority of the range inferred from the broad lines by G15), but GW-driven and decoupled if
$M>10^{8.7} {\Msun}$. However, for the shorter binary periods $3 \lsim
\chi\lsim8$, justified for $0.3 \lsim q \lsim 0.8$, we find
that the inferred smaller binary separation would place the binary
well past decoupling.  For $q>0.3$ and $\alpha<0.3$, the binary in
PG~1302 is plunged even deeper into the GW--dominated regime.

Because the binary outpaces the disc, it has
been argued that the post-decoupling BHs may be ``starved'' and thus
dim~\citep{Milos:Phinney:2005,Shapiro:2010,TM10}. Recent simulations
\citep{Noble+2012, Farris:2015:GW} show that high levels of accretion
can persist well past the decoupling phase, delivering gas to the
binary efficiently until much closer to coalescence. These simulations
also exhibit the lopsided cavity which generates the $\chi t_{\bin}$
variability considered here. Identification of the variability in PG
1302 with the cavity wall lump period would constitute the (to our
knowledge, first-ever) detection of an SMBHB which is undergoing GW
dominated inspiral, yet producing bright emission, near the Eddington
limit.\vspace{-1\baselineskip}\footnote{Note that in the precessing
  binary model for OJ287 \citep{LehtoValtonen1996,Valtonen+2008}, the
  orbital period is 12.2 yr, the primary is very massive ($\sim
  1.8\times10^{10}~{\rm M_\odot}$), but the secondary is light ($\sim
  1.4\times10^{8}~{\rm M_\odot}$). The latter reduces the efficiency
  of GWs, but increases the impact of a gas disc; as a result, the
  OJ287 binary is gas-driven, well before decoupling.}

\subsection{Binary Fraction among Quasars}
\label{SS:Rates}

A shorter binary orbit also reduces the expected number of detectable
SMBHBs in quasars, because gas driven binaries are expected to spend less time at
smaller separations.  A simple estimate of the fraction of quasars
that would harbour binaries with an orbital period $t_{\rm bin}$ can be
obtained from the residence time $t_{\rm res} = a/ \dot{a}$ at each
separation $a$, and the lifetime of bright ($L_Q /L_{\rm{edd}} \gsim
0.3$) quasars, $t_Q \sim 10^8$ yr \citep{PMartini:2004}.  Assuming
that a fraction $f_{\bin}$ of all quasars are triggered by coalescing
SMBHBs (e.g \citealt{Hopkins2007a} and references therein), it follows
that among bright quasars, the fraction with orbital period $t_{\bin}$
is $f_{\rm var}=f_{\rm bin}f_{\rm duty}$, where $f_{\rm{duty}} = {t_{
    \rm{res}} (t_{\bin}) }/{t_Q}$ is the fraction of the bright quasar
phase that a typical binary quasar spends at the orbital period
$t_{\rm bin}$.

We use the binary+disc models of HKM09 to predict the residence times
of binaries.  Prior to decoupling, $t_{\rm res}$ is determined by the
binary's interaction with the gas disc. For the masses and separations
relevant for PG~1302, the disc would be radiation pressure dominated,
yielding a relatively shallow power-law dependence $t_{\rm res}\propto
t_{\bin}^{\beta}$ with $0.5\lsim\beta\lsim 1.5$.  These scalings
depend on the poorly understood physical model of the disc and its
coupling to the binary. Past decoupling, the residence time is
precisely known, since it is determined by the strength of GWs. The
dependence is much steeper, $t_{\rm{res}}\propto t_{\bin}^{8/3}\propto
\chi^{-8/3}$.  For reference, a binary with $M=10^{8.5}{\rm M_\odot}$
and $t_{\rm bin}=4{\rm yr}$ would be in the disc-driven stage and
would have $t_{\rm res}\approx 10^6$ yr, yielding a large $f_{\rm
  duty}\approx 10^{-2}$.

\begin{figure}
\begin{center}
\includegraphics[scale=0.23]{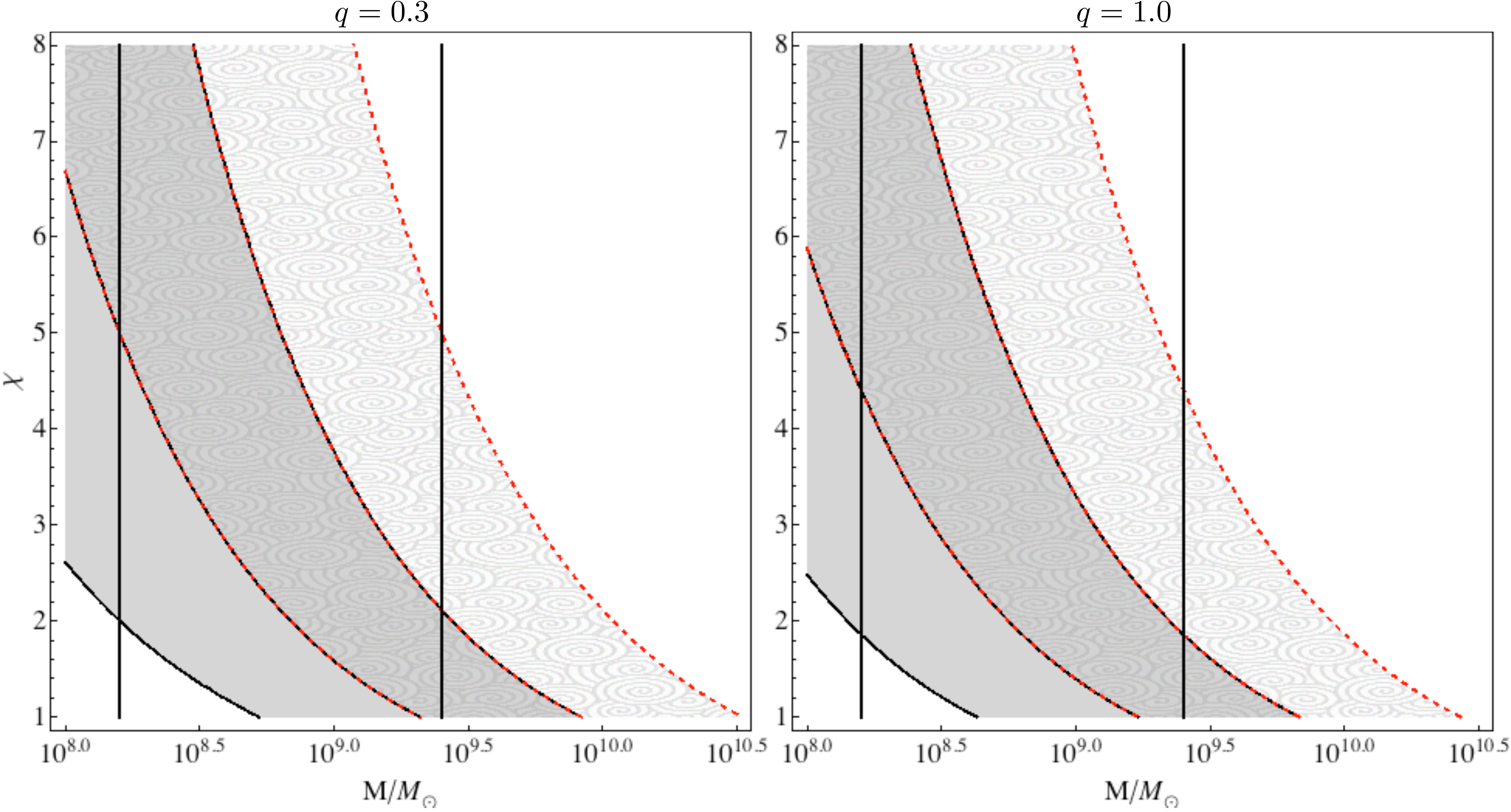} \vspace{-15 pt}
\end{center}
\caption{Combinations of total binary mass $M$ and $\chi = t_{\rm
    opt}/t_{\bin}$ for which the predicted binary fraction of quasars
  matches $10$ CRTS candidates with luminosity above that of PG~1302
  (dark grey, bounded by solid curves), and for which it consists only
  of PG 1302 (light grey, bounded by red dashed curves). Vertical lines
  delineate the range of masses preferred by broad line widths. Each
  shaded region is bounded by the fraction $f_{\bin}$ of
  quasars which are triggered by a binary. In each case, the
  lines correspond to $f_{\bin} = 0.01,0.1,1$ (left to right).}
\label{Fig:Rates}
\end{figure}

The expected $f_{\rm var}$ can be compared with the number of periodic
candidates uncovered in CRTS \citep{CRTS1:Drake:2009,
  CRTS4:Djorgovski:2011, CRTS3:2011Mahabal}.\footnote{Since the
  expected $f_{\rm var}(M,t_{\rm bin})$ declines steeply with
  increasing $M$ and decreasing $t_{\rm bin}$, it is a good proxy for
  the fraction of quasars with period $t_{\rm bin}$ {\em or less}, and
  BH mass $M$ {\em or higher} (or equivalently luminosity $L$ or
  higher, further assuming a monotonic relation between $L$ and $M$).}
There are $\approx114,000$ quasars in the CRTS sample with luminosity
higher than PG 1302, $\approx 6$ of which are SMBHB candidates with
period $t_{\rm obs}\lsim 5$yr (Graham, private communication),
amounting to an observed fraction of $f_{\var}^{\rm
  obs}=5\times10^{-5}$. Fig.~\ref{Fig:Rates} illustrates combinations
of $M$, $\chi$, and $f_{\rm bin}$, for which we expect $1$ (light grey
regions) or $10$ (dark grey) candidates in the CRTS quasar sample with
periods $\leq5.2$ yr.  Each shaded region is bounded by the assumed
fraction of quasars related to SMBHBs at all, $f_{\bin}=0.01$ (left)
to $f_{\bin}=1$ (right).

If the observed period of PG 1302 is assumed to be the binary orbital
period ($\chi=1$), then the rarity of the binary candidates in CRTS
require $f_{\bin} < 0.14$ ($<0.19$) at $q=0.3$ ($q=1.0$), even at the
most extreme mass of $M=10^{9.4} {\Msun}$. Taking the G15 fiducial
mass of $M=10^{8.5} {\Msun}$, these fractions must be as low as 
$f_{\bin} < 0.006$. These low values would be surprising, as a large fraction of quasars
are commonly believed to be triggered by mergers. This association is
based on various pieces of observational evidence, as well as on the
success of merger-based quasar population models to reproduce many
properties of the observed quasar population (e.g. \citealt{KH2000}).
If, instead, the observed period of PG 1302 is due to the $3-8$ times
longer lump-periodicity, then the SMBHB fraction and the inferred
binary mass of PG 1302-12 come into wider agreement with the
expectation that $f_{\rm bin}=O(1)$, e.g. allowing $f_{\bin} \sim 0.3$
with $q=0.3$ and $M=10^{8.5} {\Msun}$.

\subsection{Detectability of GWs} 
\label{SS:GWs}

A reduced orbital period increases the frequency and amplitude of
GWs emitted by a binary, and it is interesting
to ask whether PG~1302 may be detectable by present or future PTAs. The GW frequency, $f_{\rm{GW}}  = 2 t^{-1}_{\bin}
\approx 61 \left( \chi/5 \right)$ nHz, places the binary in the range
of PTA sensitivity (e.g. \citealt{iPTA}). We calculate an SNR for the PG 1302 binary
choosing optimistic binary parameters $M=10^{9.4} \Msun$, $q=0.5$, and
$t_{\bin} = 1447\chi^{-1}$ d. 
The GW induced rms timing residual (for simplicity, adopting the sky and
polarization averaged values) is 
$\delta t_{\rm{GW}}  = \sqrt{8/15} h/(2 \upi f_{\rm{GW}} ) \sqrt{f_{\rm{GW}}  T} \approx 2.6 \left( \chi/5 \right)^{1/6} \sqrt{T_{\rm yr}}$ ns 
for a one year observation time $T_{\rm yr}$; for
$f_{\rm{GW}} \gsim10$ nHz, the timing residual noise is nearly constant in
frequency. Using noise curves of currently operating PTAs, the SNR for
GW detection of the PG~1302 binary is $\sim 0.005 \left(\chi / 5
\right)^{1/6}  \sqrt{T_{\rm yr}}$ for NANOGrav (fig. 12 in \citealt{NANOGrav2014:SC}) or
$\sim 0.011 \left( \chi / 5 \right)^{1/6}  \sqrt{T_{\rm yr}}$ for the PPTA (black curve in
fig. 9 of \citealt{PPTA:2014:SC}). The reduced binary period increases
the SNR, but this increase is unfortunately modest. Future detectors,
such as the international pulsar timing array (iPTA;
\citealt{IPTA:2013:OP}) as well as inclusion of the square kilometre
array (SKA; \citealt{SKA:2009:OP}) in the PTA telescope networks will
improve the SNR by about an order of magnitude, but PG~1302 will
remain a factor of $\sim10$ below detection.

\section{Testing the Binary BH Scenario for PG 1302}
\label{S:Predictions}
\subsection{Broad Line Variability and Asymmetry}
\label{SS:BLs}

\cite{Jackson:1992:PGHbeta} report 
a $\sim (150\pm 50)$km s$^{-1}$ offset between the broad and narrow
line components of PG1302's H$\beta$ emission line. This is much
smaller than the secondary BH's orbital
speed,\\ $v_2=14,500~(1.5/[1+q])(M/10^{8.5}M_\odot)^{1/3}(\chi/5)^{1/3}{\rm
  km~s^{-1}}$, or the width of the broad lines. Such larger offsets
have been predicted for binary SMBHs, assuming that the broad line
region (BLR) originates from gas bound to one component of a binary
SMBH and thus shares its overall orbital motion (e.g
\citealt{Tsalmantza:2011}).  Here we argue that the smaller offset for
PG1302 can also be attributed to a binary SMBH, assuming that the BLR
is located farther out, in the circumbinary gas.  Using simple toy
models, we show that the lopsided geometry of the CBD gas
(Fig. \ref{Fig:DensLS}) could generate the small observed offset but
large width of PG 1302's broad H$\beta$ line.  The models below are
meant to be mere illustrations; a self-consistent description of the
BLR is left to future work.

The idea is that the large width of a line can reflect the orbital
speed of gas in the CBD (over a range of annuli), whereas the offset
of the line centroid is caused only by departures from axisymmetry and
can be much smaller. (In a strictly axisymmetric BLR, the blue-- and
redshifts from gas on opposing sides of the binary would be the same and
leave no net offset).  To illustrate this, we compute the line offset $V_0$ as the emission--weighted line--of--sight (l.o.s.) velocity,
\begin{equation}
V_0= \frac{ \int^{2 \upi}_0 \int_{\mathcal{R}}{\rho^{n}(v_{\phi}/r)^{m}  \ v_{\rm{los}} \ r dr d\phi}  }{ \int^{2 \upi}_0 \int_{\mathcal{R}}\rho^{n}(v_{\phi}/r)^{m} \ r dr d\phi}
\label{Eq:cent}
\end{equation}
and the width $\Gamma$ as the weighted rms l.o.s. velocity
\begin{equation}
\left(\frac{\Gamma}{ 2 \sqrt{2 \ln{2}}}\right)^2 =\frac{  \int^{2 \upi}_0 \int_{\mathcal{R}}{\rho^{n}(v_{\phi}/r)^{m} \  (v_{\rm{los}} - V_0)^2 \ r dr d\phi}  }{ \int^{2 \upi}_0 \int_{\mathcal{R}} \rho^{n}(v_{\phi}/r)^{m}  \ r dr d\phi}
\label{Eq:FWHM}
\end{equation}
over a surface patch $\mathcal{R}$ in a $q=0.5$ simulation (see
Fig. \ref{Fig:DensLS}).

In order to select the scaling indices $n$ and $m$, and to identify a
patch $\mathcal{R}$ corresponding to the BLR, we first consider a
steady thin disc model for the CBD (as in \S \ref{S:Implications}),
and assume that the CBD is illuminated by a central ionizing source
(i.e. the minidiscs).  For PG1302's parameters, the
inner region of the CBD would have density $n\sim 10^{12-13}{\rm
  cm^{-3}}$ and would be highly opaque to ionizing radiation.  In this
case, recombinations in a volume corresponding to a very thin ($\Delta
R \ll R$) inner annulus of the CBD would balance the central ionizing
photon rate.
In particular, the disc would absorb the covering fraction $2 \upi R
H/(4 \upi R^2) = 0.05 H/(0.1R)$ of the central luminosity.  PG~1302's
bolometric luminosity is $\sim 6 \times 10^{46}$ erg s$^{-1}$;
assuming that $\gsim 1/3$rd of this is emitted in the UV, $\gsim 5\%$
of which is absorbed by the CBD (i.e. $H/R\gsim 0.1$), this would be
sufficient to provide the total power $\sim 10^{45}$ erg s$^{-1}$
measured in the broad lines \citep{WangHo:2003}.  The line emission
from each patch of the CBD depends only upon the number of ionizing
photons incident on the CBD in that direction, i.e. proportional to
the scale height $H$ of the inner wall of the CBD. For an adiabatic
scale height, assuming vertical disc hydrostatic equilibrium, $n=1/3$
and $m=-1$. To be specific, $\mathcal{R}$ is chosen by excising the
binary plus minidiscs, and imposing a surface density range
$\Sigma/\Sigma_0 = (0.01, 0.5)$. While somewhat ad hoc, we find that
this $\Sigma$ range accurately picks out the streams edges inside the
cavity, and the thin inner edges of the CBD.

Although a standard Shakura-Sunyaev CBD is optically thick outside of
the inner cavity, we also consider, for generality, an alternative
scenario, where the BLR emission is produced by an optically thin
medium, but still resembling the lopsided geometry in our
simulations. In this case, line emission would scale with the
recombination rate, with $n=2$ and $m=0$ in equations.~\ref{Eq:FWHM}. We
adopt the region $\mathcal{R}$ to be an annulus with inner and outer
radii $(2a, 6a)$, chosen to encompass the inner CBD.

\begin{figure}
\begin{center}
\includegraphics[scale=0.30]{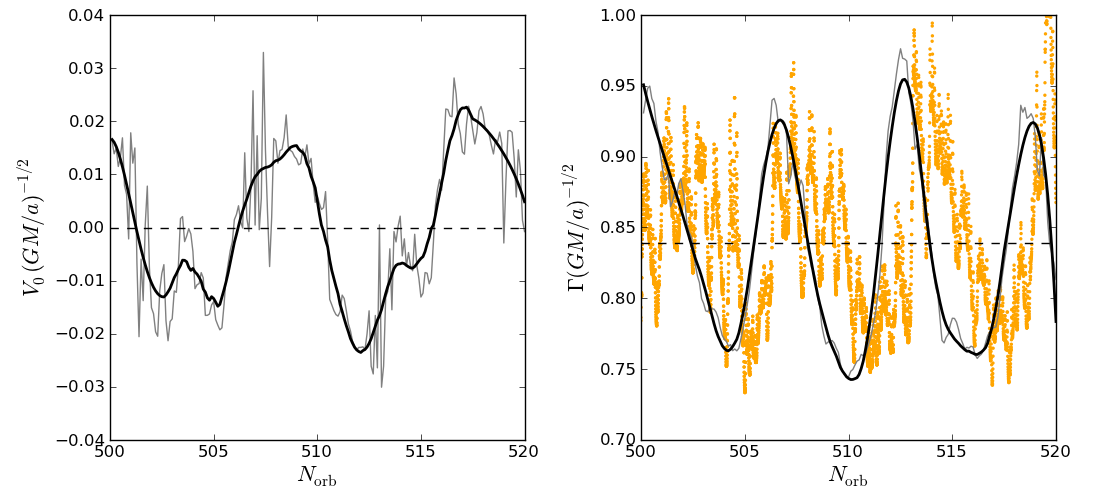}  \vspace{-25 pt}
\end{center}
\caption{Predicted variations of the centroid $V_0$ (left) and FWHM
  $\Gamma$ (right) of an emission line emanating from the inner 
  CBD. The total accretion rate on to both black
  holes is over plotted in the right-hand panel in arbitrary units
  (orange). Dark black lines are smoothed versions of the light grey
  simulation data.}
\label{Fig:BLs}
\end{figure}

We use the simulated surface density $\Sigma$, azimuthal velocity
$v_{\phi}$ and the inferred isothermal scale height of the disc $H =
rc_{\rm{s}}/v_{\phi}$ (assuming vertical hydrostatic equilibrium), to compute
the volume density $\rho = \Sigma/H$ and l.o.s. velocity
$v_{\rm{los}} = v_{\phi} \cos{\phi}$.  All l.o.s. velocities are
multiplied by an additional factor of $\sin i$, where $i$ is the CBD
inclination angle, measured from face-on.  We calculate $V_0$ and
$\Gamma$ 10 times per orbit for 20 orbits.  Fig. \ref{Fig:BLs}
displays the variations of $V_0$ and $\Gamma$ with time for the
optically thin case.  The line centroid varies with mean and range
\[
\begin{array}{l}
 V^{\mbox{thn}}_0=(\ \ \  2 \pm 167) \left[\frac{\sin i}{\sin (14.1^{\circ}) }\right] \\
 V^{\mbox{thk}}_0=(-5 \pm 285)  \left[\frac{\sin i}{\sin (11.4^{\circ}) }\right]
\end{array}
\frac{\rm km}{\rm s}
\left(\frac{M}{10^{8.5} \Msun}\right)^{\frac{1}{2}}
\left(\frac{a}{94 R_{\rm{S}} }\right)^{-\frac{1}{2}},
\]
 while the line full width at half-maximum (FWHM) fluctuates periodically on the lump's orbital time,
 \[
\begin{array}{l}
 \Gamma^{\mbox{thn}}  =  (4450 \pm 615)  \left[\frac{\sin i}{\sin (14.1^{\circ}) }\right] \\
\Gamma^{\mbox{thk}}  =  (4450 \pm 636)   \left[\frac{\sin i}{\sin (11.4^{\circ}) }\right]
\end{array}
\frac{\rm km}{\rm s}
\left(\frac{M}{10^{8.5} \Msun}\right)^{\frac{1}{2}}
\left(\frac{a}{94 R_{\rm{S}} }\right)^{-\frac{1}{2}}.
\]
The fiducial inclination angles above are chosen to match the observed H$\beta$ FWHM.

We find that both models of a CBD BLR predict a
line offset consistent with that observed in
\cite{Jackson:1992:PGHbeta} and which require a CBD inclination angle
that also predicts consistent line widths. Additionally, we find that
$\Gamma$ varies by up to $\sim 14$ per cent of the mean in each case.
It is important to emphasize that in the optically thick case, these
results arise from the non-axisymmetric velocities in the gas that
trace the lopsided inner wall of the CBD, whereas in the
optically thin case, they are driven by the lopsided density
distribution.

Since fluctuations in the latter case arise from the lump's varying
position along the cavity wall, they correlate with long--term
variations in the BH accretion rate. In the right-hand panel of
Fig. \ref{Fig:BLs}, we plot the accretion rate on to the BHs, together
with the $\Gamma$ variations of the optically thin case. The phase lag
between line-width maximum and accretion maximum derives from the time
between lump passage near the BHs, and the lump--enhanced accretion in
the minidiscs. The phase difference is therefore independent of the
observer's viewing angle. This is not true for the amplitude and shape
of the FWHM variations, and the relative phase of $V_0(t)$, which
depend on viewing angle. For the optically thick case, the FWHM
variations are similar, except that they have a $\sim20\%$ higher
mean, and are $\sim$half a cycle out of phase with the accretion rate
modulation. The phase difference arises because the optically thick
model tracks the low density gas at the cavity and stream edges
instead of the high density lump. The centroid variations are also
similar in the optically thick case but have a few times higher mean
and deviation because they track the eccentric cavity shape rather
than a symmetric annulus.

The line characteristics which we calculate here are dependent on the
existence and magnitude of the orbiting cavity wall over-density
(requiring $3 \lsim \chi \lsim 8$). As long as the binary has mass
ratio such that a cavity wall lump is generated, our BLR calculation
is largely unchanged. In addition to mass ratio, disc viscosity and
temperature affect the lump size and thus the magnitude of broad line
variations. The line shape also depends on the broad line emission
model. Hence, a full study of CBD broad lines, which examines more
sophisticated recombination models and a range of disc parameters,
binary mass ratios and viewing angles, is warranted in a future study.
Parameter dependences aside, observation of line variability, matched
to luminosity variability, would provide evidence for the CBD model
and the origin of the BLR as well as identification of the CBD cavity
wall period.

\subsection{Relativistic Effects}
\label{SS:Rel}

{\em Beaming.} 
\citet{DHS+2015} have shown that if PG1302 consists of
a massive ($M\gsim 10^{9} \Msun$) but unequal-mass ($0.03 \lsim
M_2/M_1 \lsim 0.1$) SMBH binary, seen within $\lsim 30^{\circ}$ of
edge-on, then the entire 0.14 mag variability of PG 1302 can be
explained by relativistic Doppler boost.  In the hydrodynamical
explanation proposed here, where the 5.2-yr modulation arises from
variations in the accretion rate, relativistic boost would also
inevitably imprint additional sinusoidal modulations at the true
(shorter) binary period. The effect would be enhanced, because the
secondary's velocity is higher by a factor of $\chi^{1/3}$,
potentially causing a detectable second peak in the periodogram at
$5.2\chi^{-1}$ yr.  Requiring consistency with \S~\ref{SS:BLs}, we
use the maximum binary mass and minimum mass ratio ($q=0.03$) to put an
upper limit on the secondary's l.o.s. velocity $v_{\rm{los}}$.
The relativistic beaming factor is
$[\Gamma(1-v_{\rm{los}}/c)]^{\alpha-3}$, where $\Gamma$ is the Lorentz
factor. \citet{DHS+2015} have estimated the spectral index $\alpha=1.1$
from an average over the continuum in the $V$ band.  We find that the
corresponding maximum velocity imprints a 0.07 mag amplitude
modulation on PG 1302's light curve. PG~1302's periodogram does not
show a significant secondary peak with sub-5.2 yr periods, but noise
modelling suggests that such second peaks would be detectable only at
amplitudes of $\gsim0.07$mag, $\sim$ half of the 5.2-yr modulation
\citep[][see the next section]{Charisi:2015:PG1302}.

\noindent{\em Iron K $\alpha$ lines}. Because the binary separation can
be reduced below $\lsim 100 R_{\rm{S}}$, FeK $\alpha$ lines generated at such
small separations can have characteristic binary-related features,
such as `missing wings' (due to the central cavity), or `see-saw
oscillations' of the red and blue wings (due to Doppler-shifting of
the emission from minidiscs; \citealt{McKFeZoltan:2013}).  These may be
detectable with the upcoming Astro-H mission \citep{AstroH:2014}.

\subsection{Orbital time-scale Variability}
\label{SS:OTV}
The binary+CBD model discussed above generically predicts multiple
periodic variations. If the observed period of PG~1302 is the true
binary period, then its periodogram could contain lower frequency,
higher-amplitude, and also higher frequency, lower amplitude peaks.
These could be revealed in future data, combined with more
sophisticated search algorithms for periodicity
(e.g. \citealt{VanderPlasIvezic2015} and references therein).  It will
be helpful in such a search that two of the periodicities occur at
$t_{\bin}$ and $\approx 0.5t_{\bin}$, i.e. with a characteristic 1:2
ratio in frequency. These are expected to be the only two peaks
present for $0.05\lsim q \lsim 0.3$.  The variability at $t_{\bin}$
can disappear entirely, but this happens only in the limit of
$q\rightarrow1$ (presumably rarely realized in nature) .  Thus the
detection of a secondary peaks, and the characterization of the full
variability structure, can help confirm the binary nature of PG 1302,
and constrain its parameters.  \citet{Charisi:2015:PG1302} 
searched PG~1302's available
photometric data for the existence of additional peaks at frequencies
above or below the strongest and unambiguous $5.2$ yr period. No
significant peaks were detected, and an upper limit of $\delta m \gsim
0.07-0.14$ mag (depending on frequency) was derived for the amplitude of
additional modulations.

\vspace{-\baselineskip}
\section{Conclusions}
\label{S:Conclude}

For binaries with mass ratio in the range $0.3 \lsim q \lsim 0.8$,
hydrodynamical simulations of CBDs predict dominant
luminosity variations at $3-8$ times the binary orbital period, due to
a dense lump in the CBD (Fig. \ref{Fig:DensLS}). If the periodic
variability observed in quasar PG 1302 is identified with this lump
period, rather than the orbital period of a putative SMBHB, a two to four
times smaller binary separation is inferred. This would place the
PG~1302 binary securely in the GW-driven regime, making it the first
EM detection of such a system, and proving that gas can follow the
binary past decoupling. This is encouraging for the possibility of
locating EM counterparts of GW sources.  Because binaries spend less
time at smaller separations, a shorter $t_{\rm bin}$ is in better
agreement with the small number of SMBHB candidates reported by
G15. The higher orbital velocity of the binary increases the effects
of relativistic beaming, causing optical variability at the orbital
period, and also on inferred broad line widths.  

The binary+CBD model can be tested as it predicts variability at
multiple, well-defined frequencies which depend on binary mass ratio
and disc parameters. Since a recent search \citep{Charisi:2015:PG1302}
did not reveal secondary variability in the optical light curve of PG
1302, follow up observations are required.  Finally, associating the
BLR with the inner annuli of a lumpy CBD, we find that
the FWHM of the lines can vary at the period of the continuum variability by $\pm 14$ per cent; we also predict a much smaller shift
of the broad line centroids.  These predictions are consistent with
existing observations of the width and offset of the H$\beta$ broad
line. Follow-up spectra, sampling PG~1302's apparent $5.2$ yr
period, could test this interpretation of the BLR
and aid in identifying the nature of PG~1302's variability.

\vspace{-\baselineskip}
\section*{Acknowledgements}
Resources supporting this work were provided by the NASA High-End
Computing (HEC) Programme through the NASA Advanced
Supercomputing (NAS) Division at Ames Research Center and by
the High Performance Computing resources at Columbia University.
The authors thank Maria Charisi, Imre Bartos, Adrian Price-Whelan,
Jules Halpern, and Roman Rafikov for useful discussions. We
also thank Matthew Graham and George Djorgovski for useful information
on PG~1302, as well as for providing the light curve in electronic
form. We also thank the anonymous referee
for comments that helped to improve this Letter. We acknowledge support from a National Science Foundation
Graduate Research Fellowship under Grant no. DGE1144155 (DJD) and NASA
grant NNX11AE05G (ZH and AMF).

\vspace{-\baselineskip}
\bibliography{Dan}

\end{document}